\UseRawInputEncoding
\documentclass[twocolumn,english,prb]{revtex4-1}
\usepackage[caption=false]{subfig}
\usepackage{rotating,booktabs}
\usepackage{graphicx}
\usepackage{dcolumn}
\usepackage{bm}
\usepackage{braket}
\usepackage{physics}
\usepackage{ragged2e}
\usepackage{tabularx}

\usepackage{amsfonts}

\DeclareMathSymbol{\shortminus}{\mathbin}{AMSa}{"39}

\newcolumntype{Z}{>{\centering\let\newline\\\arraybackslash\hspace{0pt}}X}

\begin{document}

\preprint{APS/123-QED}

\title{ Wurtzite/Zincblende Crystal Phase GaAs Heterostructures in the Tight Binding Approximation}

\author{Joseph Sink}
    \email{joseph-sink@uiowa.edu}
    \affiliation{  Department of Physics and Astronomy and Optical Science and Technology Center, University of Iowa, Iowa City, Iowa 52242}

\author{Craig Pryor}%
    \affiliation{ Department of Physics and Astronomy and Optical Science and Technology Center, University of Iowa, Iowa City, Iowa 52242}
\date{\today}

\begin{abstract}
Crystal phase semiconductor heterostructures allow for electron confinement without uncertainties caused by chemical intermixing found in material heterostructures and are candidates for next generation optoelectronics devices ranging from single-photon emitters to high efficiency LEDs.
While there has been a great deal of experimental work developing fabrication processes for these structures, theoretical calculations have been limited due to a lack of atomistic models that are able to incorporate the zincblende and wurtzite within the same structure.
Here, we present calculations of the electronic energies in GaAs nanowires containing various thicknesses of zincblende and wurtzite layers using a recently developed tight-binding model for wurtzite III-V semiconductors that is compatible with a zincblende model.
By comparing results in the flat-band and the unscreened  limits, we explain the sensitivity of experimentally observed band gaps on zincblende and wurtzite well widths. Our calculations suggest that experiments on devices are likely near the flat-band limit under typical operating conditions.
\end{abstract}

\maketitle

\section{Introduction}

Polytypic semiconductor heterostructures differ from material heterostructures in that the crystal phase is varied rather than the chemical composition. This allows for high quality interfaces  \cite{Lehmann.nr.2012} with low lattice mismatch \cite{Yeh.prb.1992} and defect densities that make them interesting candidates for carrier manipulation in next generation quantum devices.
The most widely studied of such structures are III-V semiconductor nanowires, in which growth conditions are tuned to controllably \cite{Geijselaers.apl.2021} alternate between zincblende (ZB) and wurtzite (WZ) crystal structures \cite{Koguchi.jjap.1992,Joyce.nl.2010,Lehmann.nl.2013}.
Experimental investigations have demonstrated the ability to fabricate quantum wells (QW) \cite{Vainorius.nl.2015,Bavinck.nl.2016,Assali.nl.2017} and dots \cite{Akopian.nl.2010,Loitsch.nl.2015,Loitsch.njp.2016} in GaAs and InP nanowires. 
These devices are being actively investigated for their use in next generation solar cells, LEDs, single-photon emitters \cite{Akopian.nl.2010,Loitsch.nl.2015} and high speed quantum sensors \cite{Spies.iop.2019,Mauthe.NatureComm.2020}.

While there has been significant progress in the growth and characterization of these structures, theoretical investigations have so far been limited by the difficulty of combining ZB and WZ crystals in the same calculation.
Some work has been done using single-band effective mass models \cite{Vainorius.Nanoscale.18,Geijselaers.apl.2021} and 8-band $\mathbf k\cdot \mathbf p$ theory \cite{Barettin.ieeex.2013}.
Since the WZ and ZB forms of binary III-V materials differ only in their atomic positions, an atomistic model such as empirical tight-binding (TB) would seem well suited.
TB models have been widely used to model the electronic properties of ZB III-V semiconductor heterostructures \cite{Lee.prb.2001,Santoprete.prb.2003,Zielinski.prb.2010,Zielinkski.PRB.12,Cygorek.PRB.20}.
Unfortunately, the most commonly used TB models for bulk III-V materials are based on first nearest neighbor Slater-Koster TB, which is known to lack polytypic transferability \cite{Bhattacharya.IOP.2010,QUAMBO.Wang.Springer.2009}.

To address this issue, we recently \cite{Sink.Adv.2023} proposed a method of dealing with the lack of polytypic transferability in Slater-Koster TB models by extending the semi-transferability treatment for single polytype chemical variations \cite{Jancu.prb.1998,Jancu.APL.2010,Laref.PSS.08} to polytypic variations. The availability of a semi-transferable III-V TB model for WZ III-V's allows the treatment of systems with WZ-ZB crystal phase heterojunctions.

We present calculations of 1D GaAs WZ-ZB superlattices, including effects from strain and polarization, using a 20-band $spds^*$ TB model. 
Effects from strain-induced piezoelectric polarization, and the spontaneous polarization in the WZ region, were included.
We treat the resulting electrostatic potential in two limits: the metallic or flat-band limit, in which the carrier density is sufficiently large to completely screen the polarization potential and the insulating limit in which there are no free carriers, and thus the polarization potential is entirely unscreened.
Our results are applicable to intermediate radius nanowires and larger, which typically have diameters $d \gtrsim 100 ~\rm nm$ and can be treated effectively as having 1D axial confinement.

Sec.\ref{subsec:geometry} describes the polytype crystal structures and the material parameters used in this work.
Sec.\ref{subsec:TB} outlines the construction of the TB Hamiltonian with strain, and the inclusion of the polarization potential arising from the gradient in the polarization potential across the interface.
 Sec.\ref{sec:ResultsAndDiscussion} presents the results, with electronic energies as functions of WZ and ZB well widths.

\section{Background and Methods}\label{sec:backgroundAndMethods}

\subsection{Geometry: WZ and ZB Stacking}\label{subsec:geometry}

\begin{figure}
    \centering
    \includegraphics[width=.6\linewidth]{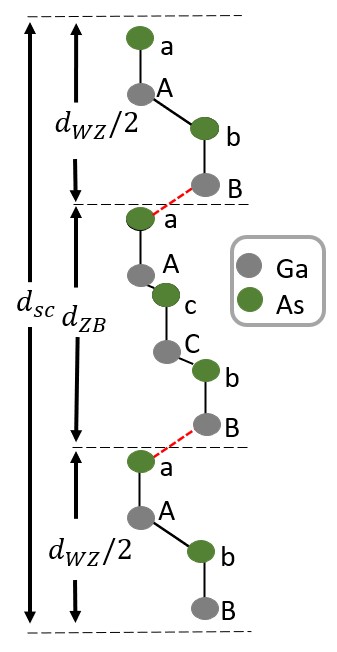}
    \caption{Polytype heterostructure SL unit cell, containing two WZ unit cells and one (non-primitive) ZB-$[111]$ unit cell. By considering the crystal environment out to second-nearest neighbors, atoms can be unambiguously categorized as either WZ or ZB. First and second-nearest neighbors of a C-layer atom are identified as ZB, whereas atoms further away are WZ. The red lines represent the NN bond that are averaged at the interface between WZ and ZB segments.}
    \label{fig:heterostructure_supercell_example}
\end{figure}

In materials with tetrahedral coordination, rotating  layers of anion-cation dimers by $\pi/3$ about specific axes maintains the global coordination number. 
This allows materials to take on a wide range of configurations, known as polytypes, that differ only in the sequence of the rotated  layers, called A, B, and C in Ramsdell notation.
Hexagonal WZ viewed along the $[0001]$-axis has AB stacking, while cubic ZB viewed along the analogous $[111]$-axis has ABC stacking. Since stacking layers preserve the local tetrahedral symmetry and chemistry regardless of the stacking order, these interfaces are exceptionally well lattice matched (Table \ref{tab:mat_const_table}). 
Here we consider  superlattices constructed from alternating 2D slabs of WZ and ZB GaAs with varying thicknesses along the $\hat{z}$-axis (Fig. \ref{fig:heterostructure_supercell_example}).

Provided each layer shares a common internal lattice constant, $u$, the atomic positions for an arbitrary stacking sequence in our supercell can be expressed simply in terms of the supercell lattice vectors,
\begin{align}
    &\mathbf{a}_1=\{\frac{1}{2},\frac{\sqrt{3}}{2} , 0\}a && \quad  \mathbf{a}_2=\{\frac{1}{2},\shortminus\frac{\sqrt{3}}{2} , 0\}a\label{eq:WZ_lat}\\
    &\mathbf{a}_3=\{0,0,c\}\nonumber
\end{align}
and basis vectors,
\begin{align}
    &\mathbf{t}_A(n)=\{0,0,n c/N\}                   &&   \mathbf{t}_a(n)=\{0,0, (2uc+n c)/N\}\nonumber\\
    &\mathbf{t}_B(n)=\{0,\frac{a}{\sqrt{3}},n c/N\}  &&   \mathbf{t}_b(n)=\{0,\frac{a}{\sqrt{3}},(2u+n)c/N\}\nonumber\\
    &\mathbf{t}_C(n)=\{0,\frac{2a}{\sqrt{3}},n c/N\} &&   \mathbf{t}_c(n)=\{0,\frac{2a}{\sqrt{3}},(2u+n )c
    /N\}\label{eq:zb_basis}
\end{align}
where $a$, $c$ and $u$ are lattice parameters, $n$ indexes the layer depth starting from n=0, $N$ counts the total number of layers in the structure, and $c=\frac{N a}{\sqrt{12u-3}}$. The upper and lower case subscripts on the basis vectors are used to differentiate the two atoms belonging to a particular Ramsdell layer (i.e., `A' and `a' are in the `A'-layer). While a common $u$-value is used to facilitate defining the initial structure, the atoms are free to move during the relaxation step as outlined in Sec.\ref{subsec:strain}. Note that each layer contains two atoms, denoted here using an upper and a lower case letter (e.g., `A' and `a' belong to the same n). The lattice constants for bulk WZ and ZB GaAs are listed in Table \ref{tab:mat_const_table}.

{\renewcommand{\arraystretch}{1.2}
\begin{table}[!h]
    \begin{tabularx}{\linewidth}{X Z Z}
        parameter&ZB&WZ\\
        \hline
        \hline
         $a~(\mathrm{\AA})$  & $3.9974 $ & $3.9855$ $^a$ \\
         $c~(\mathrm{\AA})$  & $9.7916 $  & $6.5590$ $^a$ \\
         $u$  & $0.3750$                        & $0.3746$ $^a$                       \\
        $\alpha=\alpha^\prime~(\mathrm{N/m}) $ &  $41.19$  $^b$  &  $41.19$  $^b$   \\
        $\beta  =\beta^\prime~(\mathrm{N/m})   $ &  $8.95  $  $^b$  &  $8.95  $  $^b$   \\
        $r_0~(\mathrm{\AA}) $                              & $2.4479$           &  $2.4435$            \\
        $r^\prime_0~(\mathrm{\AA}) $& $2.4479$ & $2.4568$\\
        $\cos\theta_0$&-0.3333&-0.3301\\
        $\cos\theta^\prime_0$&-0.3333&-0.3366\\
        $P_{sp} ~(\mathrm{C/m^2})$ & $0$                                               & $0.0027^c$          \\
        $e_{14} ~(\mathrm{C/m^2})$ & $-0.160$  $^d$  &             -                                        \\
        $e_{31} ~(\mathrm{C/m^2})$ &                        -                             & $0.15$ $^e$      \\
        $e_{33} ~(\mathrm{C/m^2})$ &                        -                             &  $-0.295$ $^e$  \\
        $\delta E_v~(\mathrm{meV})$  &           -                &   $125$ $^f$    \\
        \hline
        \hline
    \end{tabularx}
    \RaggedRight
    $^a$Ref. \onlinecite{Bechstedt.jpcm.2013}, $^b$Ref. \onlinecite{Martin.prb.1965}, $^c$Ref.\onlinecite{Bauer.APL.2014},$^d$Ref. \onlinecite{LandoltBornstein2002.ZBGaAs_piezo}, $^e$Ref. \onlinecite{Alekseev.PSS.2018}, $^f$Ref. \onlinecite{Geijselaers.nf.2018}
    \caption{GaAs material constants. The ZB lattice parameters ($a$, $c$, $u$) were calculated using the primitive ZB lattice constant $5.6532 ~\mathrm{\AA}$  \cite{Jancu.prb.1998}. $\alpha$ and $\beta$ are the Keating VFF bond length and angle parameters, respectively. The same elastic constants were used in the WZ and ZB regions. $\delta E_v$ is the WZ valence band maxima relative to the ZB valence band maxima ($\delta E_v =E^{wz}_{9v}-E^{zb}_{6v}$).}
    \label{tab:mat_const_table}
\end{table}
}

\subsection{Strain}\label{subsec:strain}

While small, the lattice mismatch present at the WZ-ZB interface results in a biaxial strain and a non-zero piezoelectric polarization.
In order to determine the strain profile for each of the structures investigated, it is first necessary to determine the relaxed atomic positions for each atom in the superlattice. This was accomplished using a generalized valence force field (VFF) model that allows for different equilibrium bond lengths and angles to account for the non-ideal WZ structure \cite{Camacho.pe.2010}. 
Since we are concerned with large nanowires we take the radius to infinity and model the wire as a superlattice
The system is modeled as a 1D chain of atoms with different polytypes along the chain with corresponding bulk crystal structure in the transverse directions.
In a planar superlattice the in-plane lattice constant conforms to the substrate and crystal relaxes in the growth direction to minimize the elastic energy.
For our nanowire we take the transverse lattice constant to be the weighted mean of the polytypes.

The relaxed atomic positions are determined by minimizing the total elastic energy. The contribution to the total elastic energy from atom-$i$ in the superlattice is,
\begin{align}
    &U_i=U^{c.p}_i+U^z_i\nonumber\\
    &U^{c.p.}_i=\frac{3}{16r_{0i}^2}\sum^3_{j=1}\biggl[\alpha_i(|\mathbf{r}_{ij}|^2-r_{0i}^2)^2\nonumber\\
    &\qquad\qquad +2\beta_i\sum^3_{k>j}(\mathbf{r}_{ij}\cdot\mathbf{r}_{ik}-r_{0i}^2 \cos\theta_{0i})^2\biggr]\nonumber\\
    &U^z_i=\frac{3}{16r'^{2}_{0i}}\biggl[\alpha'_i(|\mathbf{r}_{i4}|^2-r'^2_{0i})^2\nonumber\\
    &\qquad\qquad+ 2\beta'_i\sum^3_{k=1}(\mathbf{r}_{i4}\cdot\mathbf{r}_{ik}-r_{0i}r'_{0i} \cos\theta^\prime_{0i})^2\biggr]\label{eq:KVFF}
\end{align}
where $\alpha_i, \beta_i$ are the bond-distance and bond-angle elastic constants, respectively, and $r_{0i},  \cos \theta_{0i}$ are the bulk equilibrium bond lengths and angles (Table \ref{tab:mat_const_table}). The prime/unprimed notation in Eq. \ref{eq:KVFF} is used to distinguish the atom displaced along the $\hat{z}$-axis (primed) from the three co-planar atoms that lay in the plane normal to $\hat{z}$-axis (unprimed).

As we consider the polytypic interface to be atomically sharp, we separate terms into two distinct cases in calculating the contributions ($U_i$) to the total elastic energy. In the simple case where the central atom $i$ and all four nearest neighbors are located in the same region (i.e., WZ or ZB), the corresponding values listed in Table \ref{tab:mat_const_table} are used. For atoms located at the polytypic interface where some of the nearest neighbors are categorized as belonging to a different region than the central atom, the values for that pair of atoms are taken to be the average of the WZ and ZB values.

In nanowire heterostructures with comparable volumes of WZ and ZB, the relaxed lattice constant will be somewhere between that of the independent bulk phases. We simulate this effect by taking the shared lateral lattice constant, $a_\perp$, to be the weighted average of the bulk lattice constants in the two polytypes,
\begin{align}
    a_\perp=(a_{wz}N_{wz}+a_{zb}N_{zb})/(N_{zb}+N_{wz}).\label{eq:weighted_ave_axial_lattice_const}
\end{align}
This is reasonable given that we have assumed above that WZ and ZB are similarly compressible and are interested in systems with comparable amounts of WZ and ZB content over length scales small compared to the nanowire radius.

During relaxation of the atomic positions, we considered the system to be rigid in the lateral plane but free to expand along the transverse axis in order to minimize the VFF potential in Eq. \ref{eq:KVFF}. Strain was then calculated for each atom using a tetrahedral differencing method \cite{Pryor.jap.1998}. We found the strain components for semi-bulk like sheets of WZ and ZB to be of the form
\begin{align}
    &\epsilon_{\perp     }=\epsilon_{xx}=\epsilon_{yy}\nonumber\\
    &\epsilon_{\parallel}= \epsilon_{zz}\nonumber\\
    &\epsilon_{xy         }=\epsilon_{yz}=\epsilon_{zx}=0\label{eq:strain_form}
\end{align}
where $\perp$ and $\parallel$ are relative to the growth axis ($\hat{z}$). From Table \ref{tab:mat_const_table}, we can see that the smaller lateral lattice constant in WZ relative to ZB exerts a compressive strain on the ZB region (or conversely, ZB exerts a tensile strain on WZ). It is worth noting that in finite radius nanowires, Eq. \ref{eq:strain_form} holds only in the bulk region of the nanowire. In general, the terminating surface will introduce off-diagonal shear components \cite{Sundaresan.IJNM.2015}.

\subsection{Polarization}\label{subsec:polarization}

The polarization in a polytypic layer has contributions from both the spontaneous ($\mathbf{P_{sp}}$) and strain induced piezoelectric ($\mathbf{P_{piezo}}$) polarizations,
\begin{align}
    \mathbf{P}_i=\mathbf{P}_{sp,i}+\mathbf{P}_{piezo,i}.
\end{align}
In cubic materials, $\mathbf{P}_{sp}$ is zero by symmetry, although strain along the conventional $[111]$-axis may induce a piezoelectric polarization.
Hexagonal materials with $C_{6v}$ symmetry contain a polar axis that supports a non-zero $\mathbf{P_{sp}}$ directed along ($\hat{z}$) that is proportional to the deviation from the ideal crystal ($c/a-\sqrt{8/3}$) and ionicity (i.e., difference in electronegativity between the atomic species).

We include polarization effects using a continuum approximation.
The relaxed atomic positions are used to compute a local strain tensor from which we obtain the polarization in each polytope.
Due to the assumption of a large wire radius, the polarization is a (different) constant in each polytope and the polarization is computed using a bulk model.
The polarization is used to compute the potential which is then added to the TB Hamiltonian.

For a general strain tensor, the piezoelectric polarization for ZB and WZ within the continuum approximation, are
\begin{align}
    \mathbf{P}^{ZB}_{piezo}=
    \begin{pmatrix}
    -2e_{14}(\sqrt{2}\epsilon_{xy}+\epsilon_{zx})/\sqrt{3}\\
    \sqrt{2}e_{14}(-\epsilon_{xx}+\epsilon_{yy}-\sqrt{2}\epsilon_{yz})/\sqrt{3}\\  
    -e_{14}(\epsilon_{xx}+\epsilon_{yy}-2\epsilon_{zz})/\sqrt{3}\label{eq:ZB_piezo_general}
    \end{pmatrix}
\end{align}
and
\begin{align}
    \mathbf{P}^{WZ}_{piezo}=
    \begin{pmatrix}
    2e_{15}\epsilon_{zx}\\
    2e_{15}\epsilon_{yz}\\  
    e_{31}(\epsilon_{xx}+\epsilon_{yy})+e_{33}\epsilon_{zz}\label{eq:WZ_piezo_general}
    \end{pmatrix}
\end{align}
where $\epsilon_{ij}$ are the elements of the strain tensor, and $e_{14},~e_{15},~e_{31}$ and $e_{33}$ are the linear piezoelectric coefficients (Table \ref{tab:mat_const_table}). (Note that $\mathbf{P}^{ZB}_{piezo}$ has conventional $[111]$-axis aligned with the $\hat{z}$.)

Using the relations for the strain tensor described in Eq. \ref{eq:strain_form},  Eqs. \ref{eq:ZB_piezo_general} and \ref{eq:WZ_piezo_general} simplify to
\begin{align}
    \mathbf{P}^{ZB}_{piezo}=
    \begin{pmatrix}
    0\\
    0\\  
    -2e_{14}(\epsilon^{ZB}_{\perp}-\epsilon^{ZB}_{\parallel})/\sqrt{3}
    \end{pmatrix}
\end{align}
and
\begin{align}
    \mathbf{P}^{WZ}_{piezo}=
    \begin{pmatrix}
    0\\
    0\\  
    2e^{WZ}_{31}\epsilon^{WZ}_{\perp}+e^{WZ}_{33}\epsilon^{WZ}_{\parallel}
    \end{pmatrix}.
\end{align}
Ignoring surface effects,
the polarization is entirely directed along the growth axis.
The polarization at the polytypic interface is abrupt and is treated as piecewise constant in calculating the built-in potential. 
The surface bound charge ($\sigma_b$) can then be calculated by taking the difference of the net polarization across the interface 
\begin{align}
    \sigma_b=|(\mathbf{P}_{WZ}-\mathbf{P}_{ZB})\cdot\hat{z}|
\end{align}

\begin{figure}[ht!] 
	\centering
	\includegraphics[width=1\linewidth]{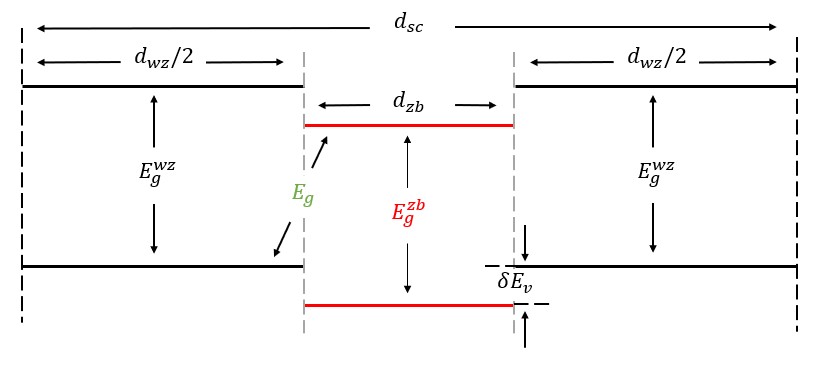}
	\caption{Schematic of supercell geometry. $d_{SC}$ is the supercell period along the growth axis, $d_{ZB}$ and $d_{WZ}$ are the widths of the ZB and WZ segments respectively. The supercell is depicted with half of the WZ polytype on either side of the ZB polytype to emphasize electron confinement. 
Both WZ and  ZB are direct gap materials. The type-II alignment results in the gap shown above, $E_g$.
}\label{fig:band_schematic}
\end{figure}

The bound sheet charge results in a spatially varying potential, generally referred to as either the built-in or polarization potential.
This additional confinement results in a quantum confined Stark shift (QCSE) which in type-II heterostructures (Fig. \ref{fig:band_schematic}) results in a well width 
dependent redshift in the PL spectrum. 
Increasing the number of free carriers available to screen the QCSE, such as by increasing the incident excitation power or current injection, results in a blueshift of the PL spectrum. 
The amount of screening present is a complicated function of the carrier concentration, which itself is generally an unknown quantity. 
To simplify this calculation, we consider  the fully insulating and fully metallic limiting regimes. 
In the insulating limit, where there are no free carriers available to screen the potential, the bound sheet charges result in triangular wells that pull apart the hole and electron wave functions to opposite sides of the heterostructure, increasing the radiative lifetimes of carriers. 
As carriers are added to the system, the bound charges are increasingly screened, resulting in progressively flatter bands. The metallic limit assumes sufficient carriers are available to fully screen out all polarization effects.

The electric field at the center (r=0) of a periodic array of infinitely wide ($R\rightarrow \infty$) 
charge sheets is
\begin{align}
    &E_z(z) =\frac{1}{2\epsilon_0\epsilon_{r}}\begin{cases}
         \sigma_b-P_{ave}                             & {\rm in~ WZ}\\
        -\sigma_b-P_{ave}                           & {\rm in~ ZB}
    \end{cases}\label{eq:internal_electric_field}
\end{align}
where $d_{ZB}$ and $d_{WZ}$ are the widths of the $ZB$ and $WZ$ regions, and
\begin{align}
    P_{ave}=\sigma_b\frac{(d_{WZ}-d_{ZB})}{(d_{WZ}+d_{ZB})}
\end{align}
is the offset required to enforce periodic boundary conditions. The dielectric constant was taken to be $\epsilon_{r}=12.88$ in both regions.

The built-in potential resulting from Eq. \ref{eq:internal_electric_field} (${\phi(z)=-\int_0^z E_z(z^\prime) dz^\prime}$) has a simple sawtooth triangular structure.
The functional form of this potential was verified by solving the Poisson equation for an axisymmetric finite radius cylindrical nanowire with spatially varying dielectric constant and polarization in the WZ, ZB and free space regions via finite element relaxation on a regularly spaced grid. It was found that the large dielectric discontinuity at the air interface efficiently suppresses the fringe field, producing a potential only weakly dependent on the radius of the cylinder. We focus on the interior of the nanowire, where the potential is linear and free from other surface effects. This is consistent with our modeling the nanowire as having an infinite radius.

\subsection{Tight Binding}\label{subsec:TB}

We use a 20-band $spds^*$ semi-transferable model for ZB \cite{Jancu.prb.1998} and WZ \cite{Sink.Adv.2023} GaAs. Far away from the interface, the unstrained system has either local $C_{6v}$ or $T_d$ symmetry, and it is trivial to identify and assign a region as bulk WZ or ZB. 
However, the region near the polytypic interface requires care, as the local symmetry is reduced to $C_{3v}$ and exists in a hybridized form of WZ and ZB. This leads to ambiguity as to which set of TB parameters should be used. To resolve this, we use a scheme of assigning atoms as belonging to either WZ or ZB based on the symmetry of the crystal environment out to second-nearest neighbors. Since both the ZB and WZ structure contain A and B layers, this criteria effectively amounts to measuring the distance of a given atom from a `C'-layer atom. In this way, the transition between the WZ and ZB crystal phase is atomically sharp. 

The onsite energies are taken to be the bulk values for their respective crystal phase without averaging. Likewise, hopping elements between atomic sites of the same polytype are set to the bulk values for that polytype. The ambiguous region is then limited to the interface bonds between WZ and ZB layers. The semi-transferable nature of the TB parameters allows for the hybridization of the bonds to be treated by taking linear combinations ($t_{interface}=t_{WZ}(1-x)+t_{ZB}x$) of the bulk WZ and ZB values. We have found that taking the mixing ratio of WZ and ZB bonds to be equal ($x=0.5$) works well without producing noticeable interface artifacts in the envelope of the TB wave function.

The valence band offset and the polarization potential, $\phi(\mathbf{r})$, are spatially varying local potentials included as offsets to the diagonal elements  of the Hamiltonian,
\begin{align}
    H^{\alpha,\alpha}_{i,i}=\epsilon^{\alpha}_{i}+\phi(\mathbf{r}_i)+(\delta E_v)_i
\end{align}
where $i$ is the atomic site index,  $\mathbf{r_i}$ is the position of atom $i$, $\alpha$ is the band index, and  $\epsilon^{\alpha}_{i}$ is the onsite energies from the TB model.

The effect of strain is included by using a power law scaling of the two-center integrals with respect to bond distance change. This method has been shown to reproduce the deformation potentials in ZB-GaAs \cite{Jancu.prb.1998}. Experimental data for the deformation potentials in WZ-GaAs are unavailable for comparison. However, the LDA calculations \cite{Wei.PRB.99} give similar pressure coefficients for WZ and ZB in III-N materials.

Note that extending this approach to smaller radii nanowires would be straight forward using a  3D model.
The strain relaxation would involve minimization of the strain energy using the conjugate gradient algorithm and the electronic Hamiltonian would be larger, though sparse.
The calculation of the strain at each atom, and thus the polarization, would proceed in the same manner as in our 1D calculation, but using relaxed atomic positions from a 3D VFF calculation.
Calculation of the piezoelectric potential would require solving Poisson's equation over the the 3D nanowire structure.

\section{Results and Discussion}\label{sec:ResultsAndDiscussion}

\begin{figure}[ht!] 
	\centering
	\includegraphics[width=1\linewidth]{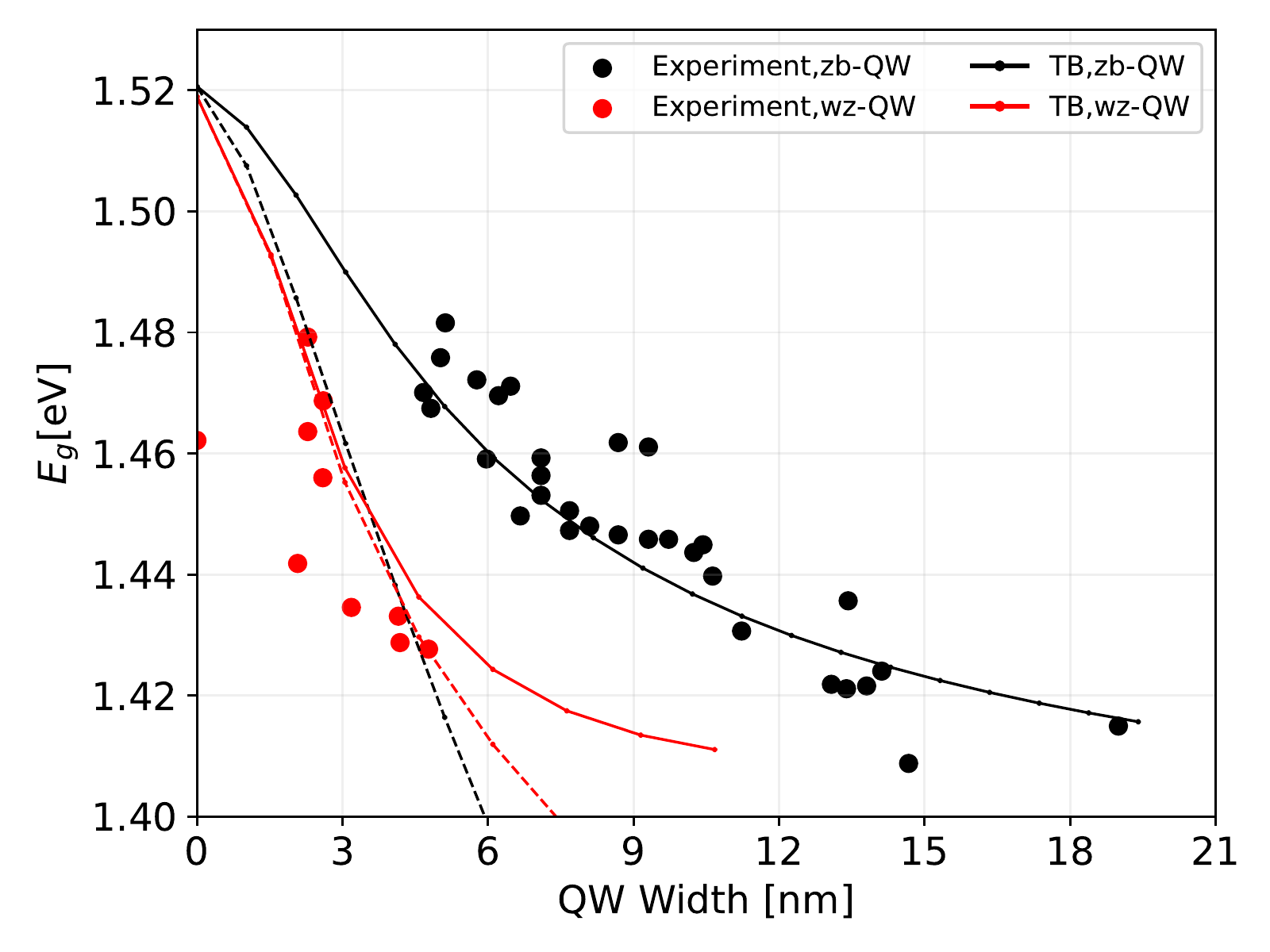}
	\caption{
	Calculated  $E_g$ as a function of   well width compared with the experimental results from Ref. \onlinecite{Vainorius.nl.2015} on single QWs in nanowires with radius $r \sim100~\mathrm{nm}$.
	To simulate a single quantum well with a superlattice calculation, the barrier width (ZB or WZ) was  fixed  to $20~\mathrm {nm}$ while the well width (WZ or ZB) was varied over a smaller range (the dot width in the figure).
	Using a barrier width of $20~\mathrm{nm}$ gave results that were insensitive to further increases in the barrier width.
	Results are shown for both the metallic (solid) and insulating (dashed) limits.
	For a WZ dot in a ZB wire (red) the calculated gap in the metallic and insulating limits are very close, and both are in good agreement with measurements.
	For a ZB dot in a WZ wire (black) only the metallic limit is in good agreement, indicating that the measurements were done under conditions of high carrier concentration.
	}\label{fig:energies_vz_well_width}
\end{figure}

\begin{figure}
\centering
\subfloat[\label{fig:gap_energies_and_sensitivity.a}]{\includegraphics[width=.45\linewidth]{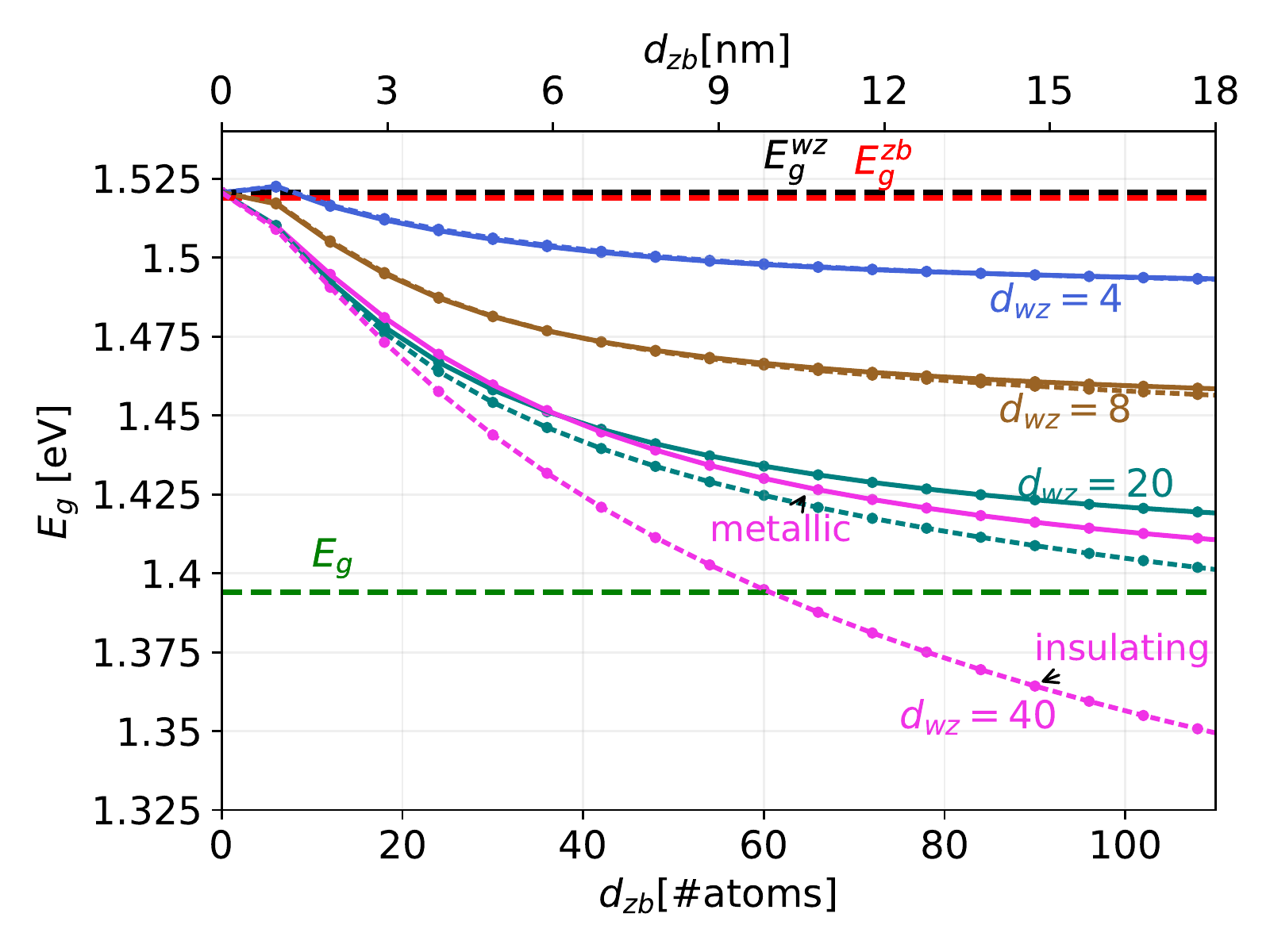}}\hspace*{\fill}
\subfloat[\label{fig:gap_energies_and_sensitivity.b}]{\includegraphics[width=.45\linewidth]{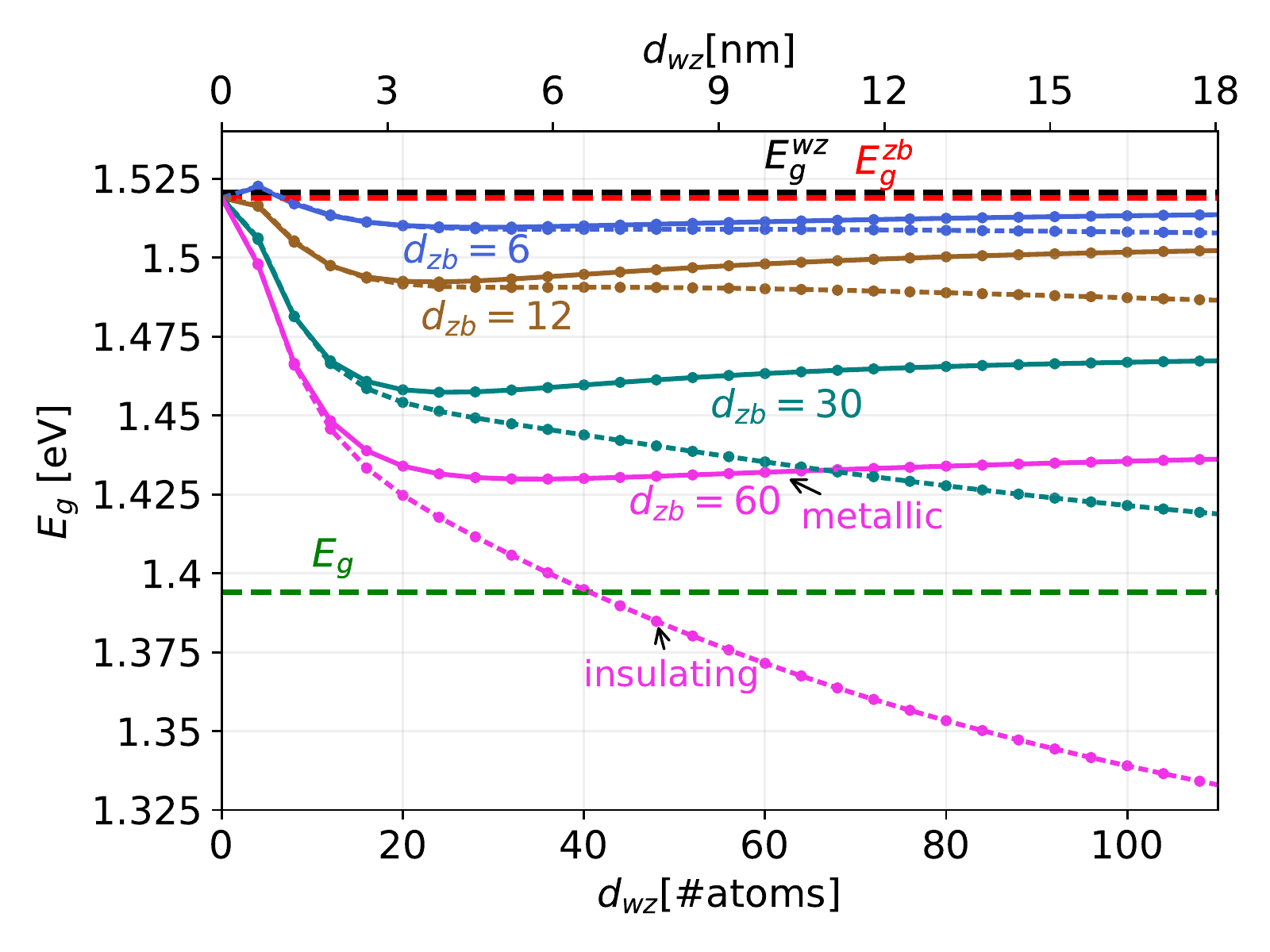}}
\\
\subfloat[\label{fig:gap_energies_and_sensitivity.c}]{\includegraphics[width=.45\linewidth]{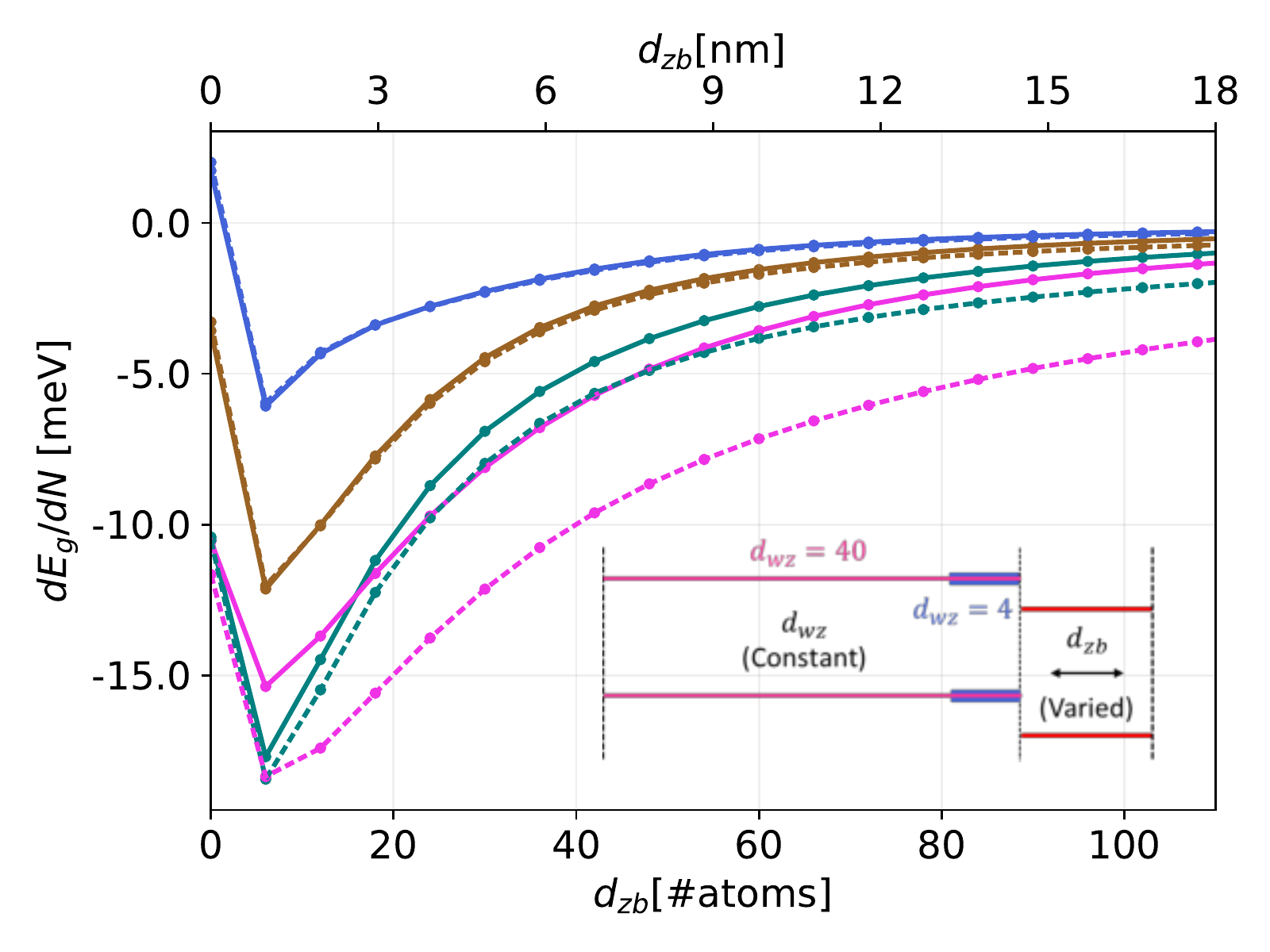}}\hspace*{\fill}
\subfloat[\label{fig:gap_energies_and_sensitivity.d}]{\includegraphics[width=.45\linewidth]{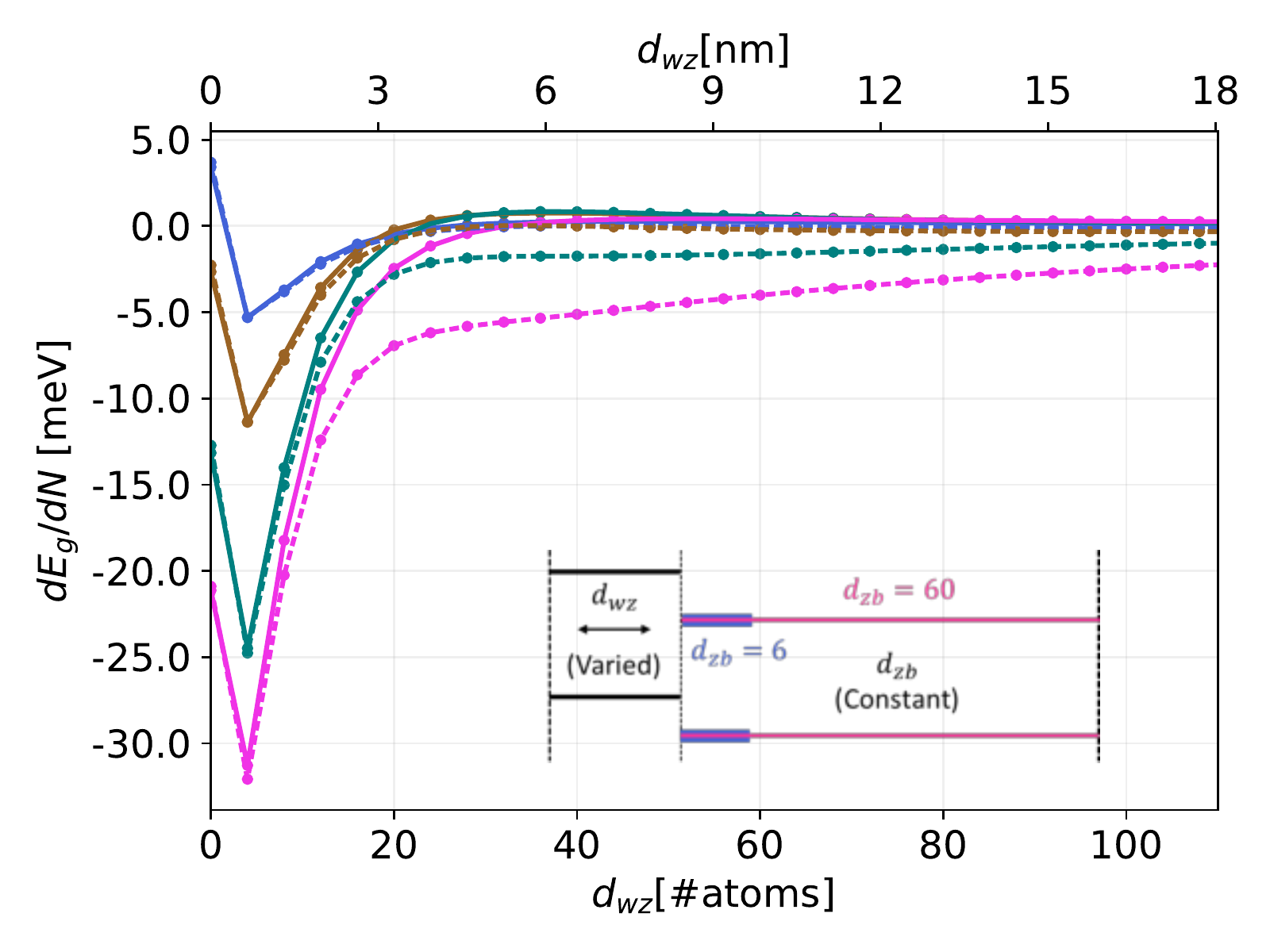}}
	\caption{
	Gap and its sensitivity to variations in layer widths.
	The upper two graphs show $E_g$ as a function of WZ and ZB widths.
	(a) $E_g$ as a function of ZB width for various fixed WZ widths (4  to 40 atoms).
	(b) $E_g$ as a function of WZ width for various fixed ZB widths (4  to 40 atoms).
	In both (a) and (b) the bulk WZ and ZB gaps are shown for comparison.
	The horizontal dashed line labeled $E_g$ is the zero confinement energy band gap limit, calculated by taking the difference of the unstrained ZB conduction band minima and WZ valence band maxima.
	The lower two graphs show the sensitivity of $E_g$ to variations in widths.
	(c) $dE_g / dN$ as a function of ZB width for various fixed WZ widths (4  to 40 atoms).
	(d) $dE_g / dN$ as a function of WZ width for various fixed ZB widths (4  to 40 atoms).
	Solid curves are for the metallic limit and dashed lines are for the insulating limit.
	}\label{fig:gap_energies_and_sensitivity}
\end{figure}

\begin{figure}[ht!] 
	\centering
	\subfloat[\label{fig:delta_energies_and_delta_sensitivity.a}]{\includegraphics[width=.45\linewidth]{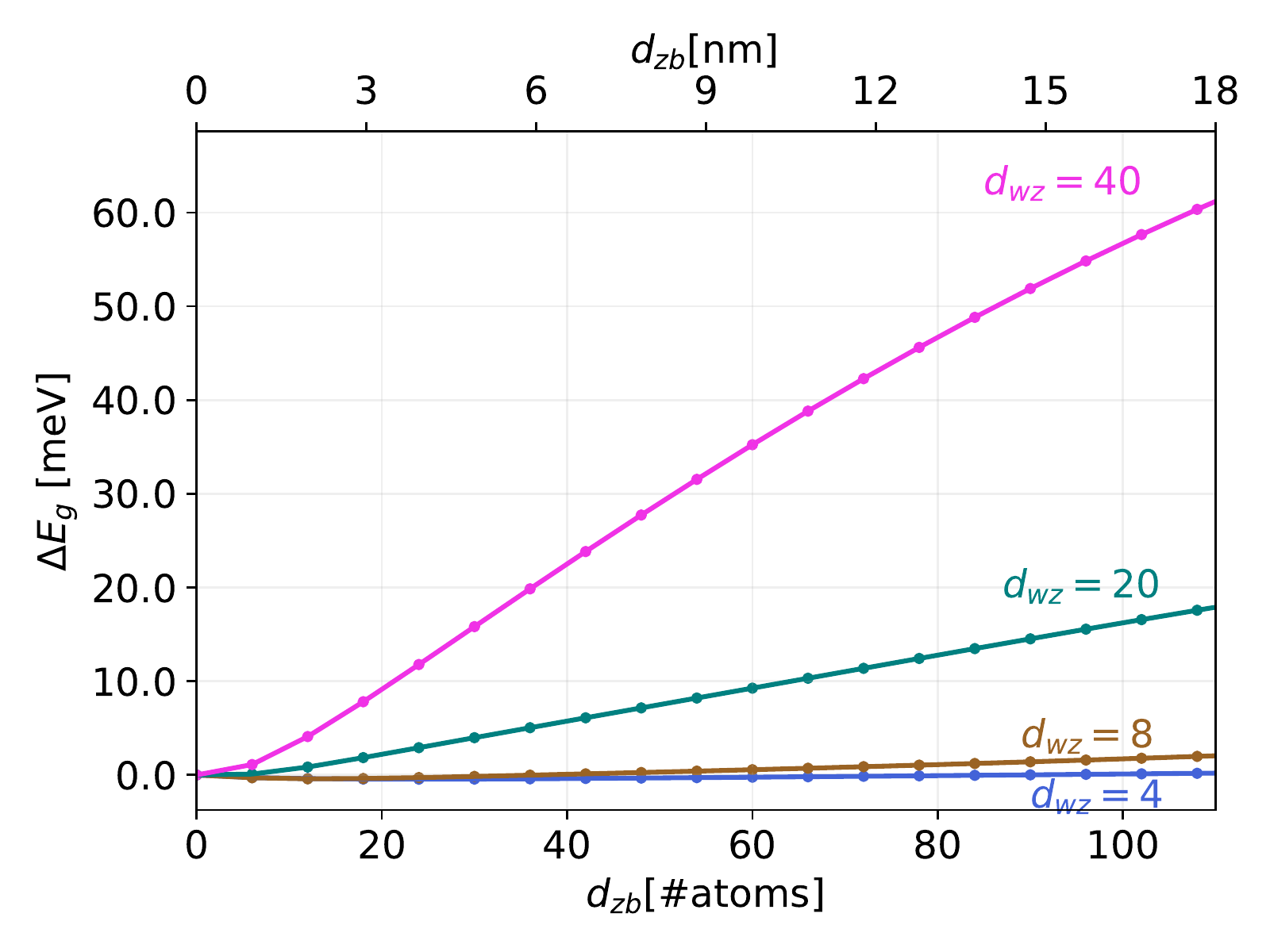}}\hspace*{\fill}
	\subfloat[\label{fig:delta_energies_and_delta_sensitivity.b}]{\includegraphics[width=.45\linewidth]{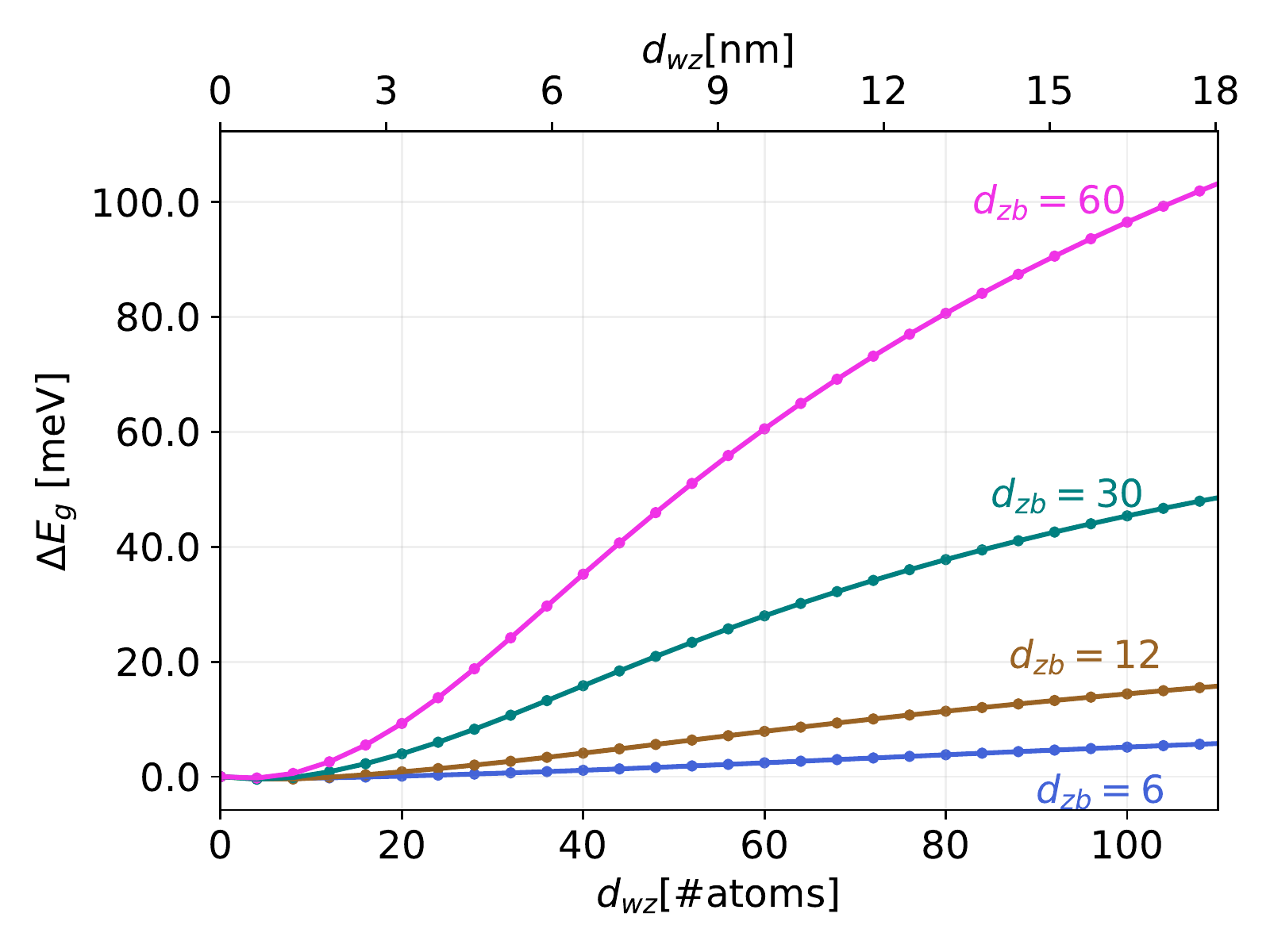}}
	\\
	\subfloat[\label{fig:delta_energies_and_delta_sensitivity.c}]{\includegraphics[width=.45\linewidth]{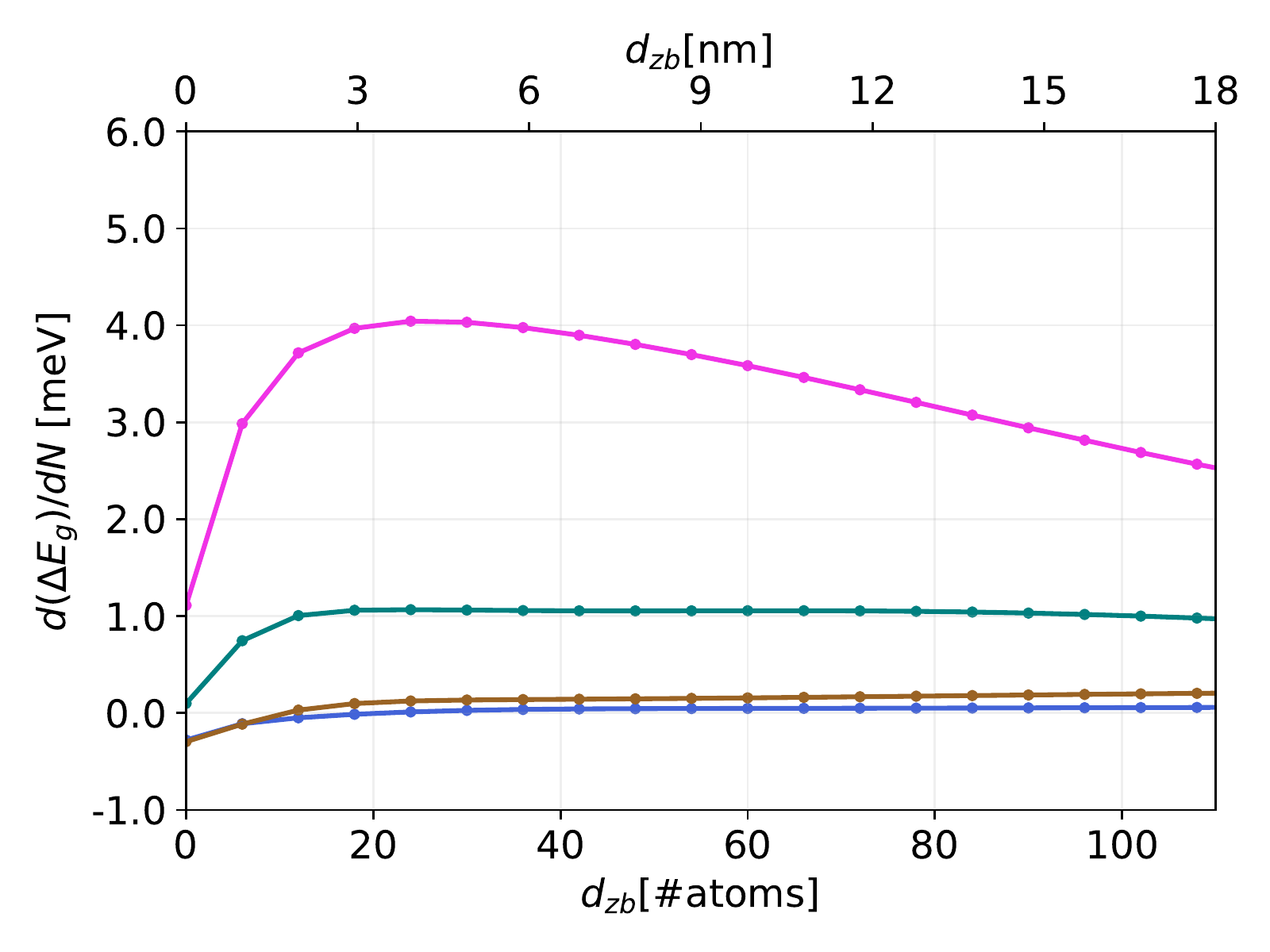}}\hspace*{\fill}
	\subfloat[\label{fig:delta_energies_and_delta_sensitivity.d}]{\includegraphics[width=.45\linewidth]{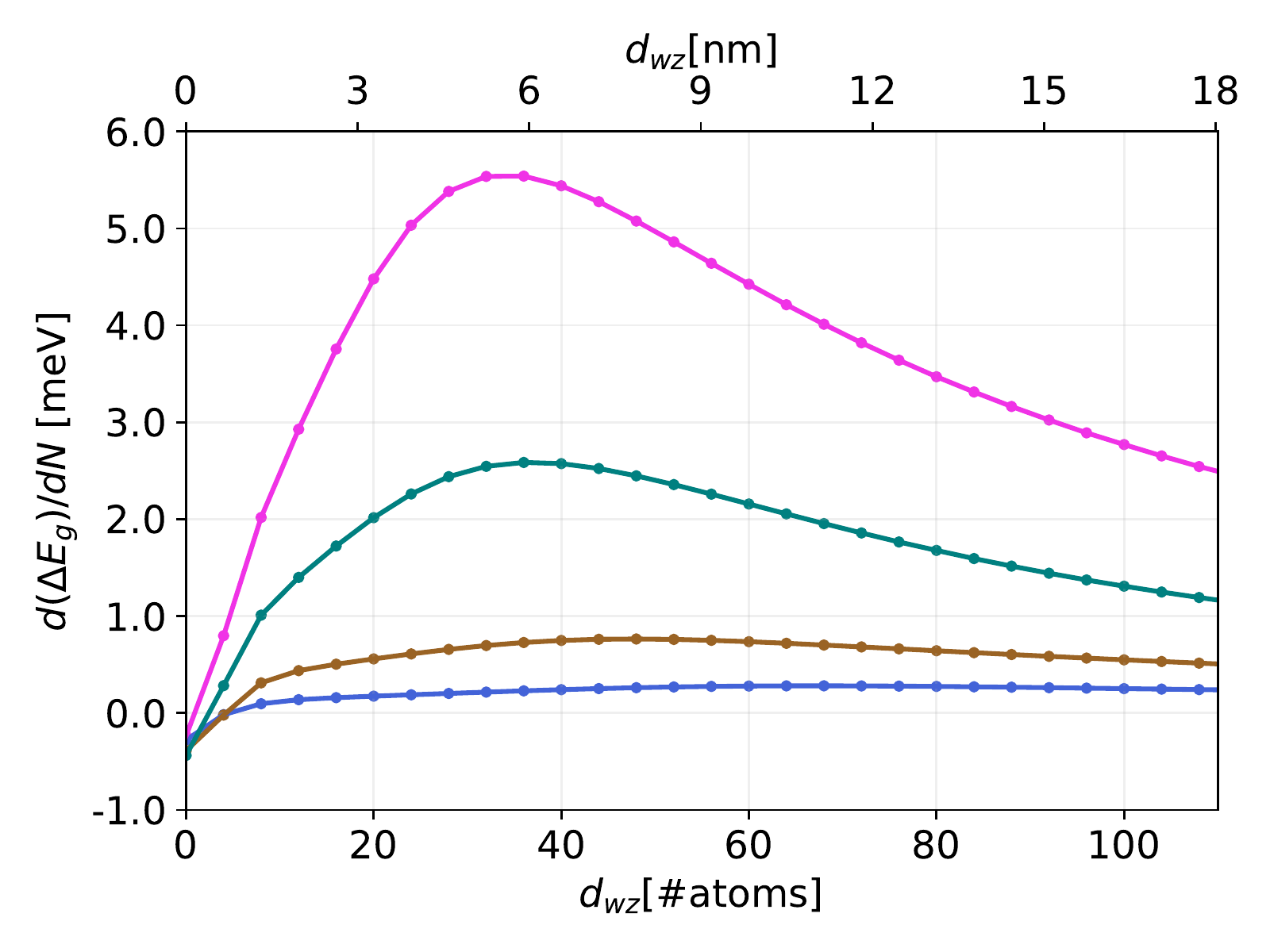}}
	\caption{
	Difference in $E_g$ between metallic and insulating limits  as a function of superlattice structure, and the sensitivity of the difference to variations in layer widths.
		The upper two graphs show $\Delta E_g= E_g^{metallic}-E_g^{insulating}$ as a function of WZ and ZB widths.
	(a) $\Delta E_g$ as a function of ZB width for various fixed WZ widths (4  to 40 atoms).
	(b) $\Delta E_g$ as a function of WZ width for various fixed ZB widths (4  to 40 atoms).
	The lower two graphs show the sensitivity of $\Delta E_g$ to variations in widths.
	(c) $d\Delta E_g / dN$ as a function of ZB width for various fixed WZ widths (4  to 40 atoms).
	(d) $d\Delta E_g / dN$ as a function of WZ width for various fixed ZB widths (4  to 40 atoms).
	Solid curves are for the metallic limit and dashed lines are for the insulating limit.
	}\label{fig:delta_energies_and_delta_sensitivity}
\end{figure}

We present direct band gap energies for a series of ZB ($d_{ZB}$) and WZ ($d_{WZ}$) widths in Figs. \ref{fig:energies_vz_well_width} and \ref{fig:gap_energies_and_sensitivity}. 
In Figs.\ref{fig:gap_energies_and_sensitivity} and \ref{fig:delta_energies_and_delta_sensitivity}, the results are split into two categories: constant $d_{WZ}$ with varying $d_{ZB}$ and constant $d_{ZB}$ with varying $d_{WZ}$.
These calculations were done with (insulating limit) and without (metallic limit) a superimposed polarization potential in order to estimate the magnitude of the expected blueshift due to increased screening from free carriers.
These calculations can be used to explain observed trends in PL spectra, as well as aid in predicting the sensitivity of the electronic states to fluctuations in carrier density and variations in WZ and ZB widths. 
While high sensitivity to structural variations is likely detrimental in many cases, sensitivity to carrier density may be beneficial in quantum sensors.

The main features in Figs.\ref{fig:gap_energies_and_sensitivity} and \ref{fig:delta_energies_and_delta_sensitivity} can be summarized as follows: a) A blueshift as screening of the built-in potential is increased (i.e., photogenerated carriers, electric gating, etc.), b) negligible QCSE for layer widths less than approximately 20 atoms (Figs.\ref{fig:delta_energies_and_delta_sensitivity.a},\ref{fig:delta_energies_and_delta_sensitivity.b})
and  c) redshift (Figs. \ref{fig:gap_energies_and_sensitivity.c},\ref{fig:gap_energies_and_sensitivity.d}) with increasing $d_{ZB}$ or $d_{WZ}$ due to decreased confinement of electrons and holes.

The small blueshift with increasing $d_{WZ}$ in Fig.\ref{fig:gap_energies_and_sensitivity.b} (solid line, metallic limit) observed after the knee at $d_{WZ}\approx 20$ atoms is caused by a decrease in strain in the WZ region from increasing WZ content (Eq. \ref{eq:weighted_ave_axial_lattice_const}) relative to the fixed ZB content. This effect is present over the whole range of structures presented, but becomes apparent only at sufficiently small confinement energies. 
In the insulating limit with large $d_{WZ}$ and $d_{ZB}$, the QCSE from the polarization potential is able to drive the gap below the bulk value.

In the flat-band/metallic limit the energy gap vs. well width (Fig.\ref{fig:energies_vz_well_width}) is in good agreement with experimental results \cite{Vainorius.nl.2015,Spirkoska.PRB.09}.
Concerning the NW-QW structures investigated by Vainorius et al. \cite{Vainorius.nl.2015}, the high sensitivity of the confined electron state in the ZB region to the polarization potential allows us to infer that the ZB QW's were operating very close to the fully screened metallic limit. Due to the large valence band mass associated with the confined hole state, the WZ-QW's show very little variation in QCSE over the range of width experimentally investigated. Additionally, we find the QCSE-shift is in qualitative agreement with analogous III-N structures \cite{Sundaresan.IJNM.2015,Wan.JAP.2001,Kazlauskas.JAP.2005,Jacopin.jap.2011} and with experimental data for GaAs  polytypic heterostructures \cite{Senichev.Nano.2018,Spirkoska.PRB.09,Vainorius.PRB.2014,Vainorius.nl.2015}. All show increasing redshifts with increasing well widths, a stronger dependence on hole confinement width (WZ region), and a blueshift with increased screening. 

\section{Conclusion}
We have used a 20-band $spds^*$ TB model to calculate the band gaps for WZbulk Zb GaAs axial polytypic heterostructure including the effects of strain and polarization. These calculations allow us to predict PL energies as well as identify structural regimes that are likely to show heightened or reduced sensitivity to sample variations in either confinement thickness or fluctuations in carrier density. This information can be used to build devices that are robust with respect to the addition/subtraction of a layer of atoms or sensitive devices that are easily modulated by external gating, current injection or photogeneration, which is a desirable characteristic in senors.

~
\onecolumngrid
~

~

~

\clearpage
\twocolumngrid
\providecommand{\noopsort}[1]{}\providecommand{\singleletter}[1]{#1}
\end{document}